\documentclass[usenatbib]{mn2e}
\usepackage[dvips]{graphicx}
\usepackage{colortbl}
\usepackage{amssymb,txfonts}

\newcommand{\apj}{{ApJ}}
\newcommand{\apjs}{{ApJS}}
\newcommand{\mnras}{{MNRAS}}

\newcommand{\nat}{{Nature}}
\newcommand{\mbh}{M_{\rm BH}}
\newcommand{\msun}{{\rm M}_{\sun}}

\def\ee{\end{equation}}
\def\be{\begin{equation}}
\newcommand{\ledd}{L_{{\rm Edd}}}
\newcommand{\mdot}{\dot M}
\newcommand{\medd}{{\dot M_{\rm Edd}}}
\newcommand{\ergs}{{\rm \,erg\,s^{-1}}}

\title[The Radiative Efficiency of Hot Accretion Flows]{The Radiative Efficiency of Hot Accretion Flows}

\author[F. G. Xie and F. Yuan]{Fu-Guo Xie,$^{1}$\thanks{E-mail: fgxie@shao.ac.cn (FGX), fyuan@shao.ac.cn (FY)}
Feng Yuan$^{1}$\footnotemark[1]\\
$^1$Key Laboratory for Research in Galaxies and Cosmology, Shanghai Astronomical Observatory, \\
Chinese Academy of Sciences,80 Nandan Road, Shanghai 200030, China\\
}

\date{Accepted 2012 August 31. Received 2012 August 27; in original form 2012 July 11}
\pagerange{\pageref{firstpage}--\pageref{lastpage}}

\begin{document}

\maketitle

\label{firstpage}

\begin{abstract}
Two significant progresses have been made in the past years on our understanding of hot accretion flows. One is that only a small fraction of accretion flow available at the outer boundary can finally falls onto the black hole while most of them is lost in outflow. Another one is that electrons may directly receive a large fraction of the viscously dissipated energy in the accretion flow, i.e, $\delta\sim 0.1-0.5$. The radiative efficiency of hot accretion flow when these two progresses are taken into account has not been systematically studied and is the subject of the present paper. We consider two regimes of hot accretion model. One is the advection dominated accretion flows (ADAFs) which lie on low accretion rate regime, $\la 10\alpha^2\ledd/c^2$; another being the luminous hot accretion flows (LHAFs) which lie above this accretion rate. For the latter, we assume that the accretion flow will has a two-phase structure above a certain accretion rate, and a simplification is adopted in our calculation of the dynamics. Our results indicate that the radiative efficiency of hot accretion flow increases with the accretion rate and is highly enhanced by the direct viscous heating to electrons compared to the previous case of $\delta\ll 1$. When the accretion rate is high, the radiative efficiency of hot accretion flow is comparable to that of the standard thin disk. Fitting formulae of radiative efficiency as a function of accretion rate for various $\delta$ values are presented.
\end{abstract}


\section{Introduction}
\label{s:intro}
One of the most important parameters in accretion theory is the radiative efficiency. This parameter describes the significance of converting rest-mass energy into radiative energy,
\begin{equation}
\epsilon\equiv {L\over\mdot c^2},\label{eq:epsilon}
\end{equation}
where $L$ is the total luminosity emitted from the accretion flow and $\mdot$ is the corresponding mass accretion rate of the system. Take the standard thin disc model \citep[][hereafter SSD]{ss73} as an example, its radiative efficiency lies in the range $0.057 - 0.43$, depending on the spin of the black hole \citep{nt73}.

According to the differences of temperature and mass accretion rate of accretion flows, we now have four accretion models which belong to two series, namely cold and hot ones. In the cold series, when the accretion rate is lower than the Eddington rate, $\dot{M}\la \medd (\equiv 10 \ledd/c^2)$, we have the standard thin disc model \citep{ss73}. When $\dot{M}\ga \medd$, the model is the slim disk \citep{a88}. The temperature of the gas in these two models is roughly within the range of $10^5-10^7{\rm K}$. The accretion flows are optically thick, emitting multi-temperature blackbody spectrum \citep{fkr02}. The radiative efficiency of the former is high, and is independent of the accretion rate \citep{nt73}. In a slim disk, the optical depth is so large that photons are trapped in the accretion flow and advected into the black hole; therefore the radiative efficiency is lower \citep{a88,m00,s09}.

In the hot series of model, the temperature of the accretion flow is almost virial. When the mass accretion rate is below a critical value, $\dot{M}_{\rm cr,ADAF}\approx 5\theta_e^{3/2}\alpha^2\medd$ with $\theta_e\equiv kT_e/m_ec^2$, we have the advection-dominated accretion flows (ADAFs; \citealt{i77,r82}, \citealt{ny94,a95}; see \citealt{nmq98,nm08} for reviews)\footnote{As we will state below, the mass accretion rate of hot accretion flows is a function of radius. In addition, the value of $\dot{M}_{\rm cr,ADAF}$ is a function of parameter $\delta$. The value of $\dot{M}_{\rm cr,ADAF}$ cited here was obtained when $\dot{M}(R)=constant$ and $\delta=10^{-3}$. The result of other cases will be presented in the present paper.}. In an ADAF the gas is tenuous. The Coulomb coupling between electrons and ions is not strong enough thus the flow is two-temperature with the ions being much hotter than the electrons \citep{ny95}. The critical accretion rate $\dot{M}_{\rm cr,ADAF}$ is determined by the balance between the Coulomb collision and viscous heating in the ions energy equation. When $\dot{M}\ll \dot{M}_{\rm cr,ADAF}$, most of the viscously liberated energy is stored as the gas internal energy and advected into the black hole rather than being transferred from ions to electrons and radiated away; therefore, the radiative efficiency of an ADAF is very low. With the increase of $\dot{M}$, more and more viscously dissipated energy will be transferred into electrons and radiated away until $\dot{M}_{\rm cr,ADAF}$ is reached at which advection is no longer dominated. ADAFs have been widely applied to the low-luminosity black hole sources including the supermassive black hole in our Galactic center, Sgr A*, low-luminosity AGNs, and the quiescent and hard states of black hole X-ray binaries \citep{yqn03,n05,y07,nm08}. When $\dot{M}\ga \dot{M}_{\rm cr,ADAF}$, Coulomb collision cooling becomes stronger than the viscous heating. \citet{y01} found that in this case, up to another critical accretion rate, there exist another hot accretion solution in which the sum of the compression work ($PdV$ work) and viscous heating balances the cooling. Compared to ADAFs, this model corresponds to higher accretion rates and radiative efficiency; thus it is called luminous hot accretion flow (LHAF; \citealt{y01}). LHAFs have been invoked to explain the origin of hard X-ray emissions detected in luminous X-ray sources such as Seyfert galaxies and luminous hard state of black hole X-ray binaries \citep{yz04,yz07}.

In the present paper we focus on the radiative efficiency of hot accretion flows. Using the data from \citet{e97}, \citet{nmq98} presented in their Fig. 7 the relationship between bolometric luminosity and the accretion rate of ADAFs (see also \citealt{ny95}). \citet{y01} investigated the radiative efficiency of LHAFs and found that $\epsilon_{\rm LHAF}$ is higher than the typical ADAF value. Both calculations are, however, based on the ``old'' version of hot accretion flow models in the sense that it is assumed that the mass accretion rate is a constant of radius and the value of parameter $\delta$, which describes the fraction of turbulent dissipation that heat the electrons directly, is very small, $\delta\sim 10^{-3}$. Both assumptions are now known to be no longer correct after the development of the accretion flow theory in the recent years, as we will illustrate in detail in \S2.

In this paper we systematically revisit the efficiency of ADAFs and LHAFs after taking into account the new progresses of hot accretion flow theory. This paper is organized as follows. We first give a brief introduction to the recent progresses on hot accretion flows in Section\ \ref{s:out_vis}. We then describe our model in Section\ \ref{s:model}. Our calculation results are presented in Section\ \ref{s:result}. The last section is devoted to a discussion.

\section{Outflow and viscous heating to electrons}
\label{s:out_vis}
Since the original work on hot accretion flows by \citet{ny94}, two major progresses have bee made, namely the existence of outflow and the importance of direct electron heating by turbulent dissipation.

\subsection{Outflow in hot accretion flows}
The evidences for the existence of outflow in hot accretion flows comes from both theoretical studies and observations. In the theoretical side, both hydrodynamical and magnetohydrodynamical simulations of hot accretion flows have found that the mass accretion rate (or more precisely the inflow rate; refer to \citealt{spb99} for details) decreases with decreasing radius, no matter the radiation is weak \citep{spb99,ia99,ia00,sp01} or strong \citep{yb10}, following a power-law scaling of radius, $\dot{M}(R)\propto R^s$. Most recently \citet{ywb12} combined all relevant numerical simulations and found that the slopes of the radial profiles of mass accretion rate obtained in various simulation works are very similar, no matter the simulations include magnetic field or not, and what kind of magnetic field configuration and initial conditions are adopted. So throughout this paper we adopt a single value for $s$. \citet{ybw12} investigated the origin of such a profile and concluded that they are because of mass loss in outflow (but see \citealt{nsp12} for a different view).

What is more exciting is that the above theoretical results have been confirmed by observations. One example is Sgr A*. {\em Chandra} observations combined with Bondi accretion theory gave the accretion rate at Bondi radius. On the other hand, radio observations put strong constrain on the accretion rate at the innermost region of the accretion flow. This rate is $\sim 1\%$ of the accretion rate at the Bondi radius. Detailed modeling by \citet{yqn03} show that $\dot{M}(R)\propto R^{0.3}$ which is close to the above-mentioned theoretical results. Another example is NGC~3115 \citep{w11}. For this source, {\em Chandra} can directly determine the density profile of the ADAF within the Bondi radius. Again it was found the result is fully consistent with the theoretical prediction.

\subsection{Direct viscous heating to electrons}
An important parameter in hot accretion flow theory is $\delta$. Early works on ADAFs have assumed that the viscous heating preferably acts on the ions, i.e., $\delta\ll 1$. However,  detailed analysis of the microphysics in accretion flow has indicated that the electrons can receive a comparable fraction of viscous heating to that of the ions \citep{bl97,b99,qg99,sq07}. For example, analytical investigation of particle heating due to Alfv\'enic turbulence by \citet{qg99} indicates that electrons can receive a significant fraction of the turbulent heating, provided the magnetic fields are not too weak ($\beta < 5 - 100$, see below for the definition of $\beta$). Similar conclusion was drawn in \citet{sq07}, who investigated the particle heating by the naturally-generated pressure anisotropy in collisionless plasma, the case of typical ADAF. Additionally, magnetic reconnection is also an important source of electron heating in the turbulent accretion flows \citep{bl97,qg99}, which can also heat the electrons significantly. Although a consensus on the value of $\delta$ has not been reached, it has been generally accepted that $\delta\sim 0.1-0.5$.

The above theoretical results have obtained observational support in the detailed modeling of Sgr A* \citep{yqn03}. In that work it was found that to explain the observations $\delta\approx 0.5$ is required. Therefore, we will focus on this value of $\delta$ in the present work. Given the theoretical uncertainties, however, a large range of the value of $\delta$ will also be considered, from $\delta=10^{-3}$ to $0.5$.

\subsection*{}
These two progresses have major impacts on hot accretion models. For example, with the existence of outflow, the compression work for ions and further the efficiency will be suppressed \citep{qn99,xy08}. Also, because of the suppression of compression work, which is a crucial heating term in LHAF equations, the $\mdot$ range of Type I LHAF (see below for definition) will be smaller compared to the original result in \citet{y01}. A larger $\delta$ obviously will increase the energy input to the electrons thus an increase of the radiative efficiency.

\section{Hot Accretion Flows: Model Description of ADAF and LHAF}
\label{s:model}

ADAFs exist only below $\dot{M}_{\rm cr,ADAF}$. Above $\dot{M}_{\rm cr,ADAF}$, the flow will enter into the LHAF regime. LHAF is thermally unstable \citep{y03}, but for accretion rate lower than another critical value, $\mdot_{\rm cr,LHAF}$, the growth timescale of the instability is longer than the accretion timescale, thus the flow can remain hot throughout the radius \citep{y03}. Above $\mdot_{\rm cr,LHAF}$, the one-dimensional steady calculations show that the radiative cooling is so strong at small radii that the flow collapses and forms a thin disk \citep{y01}. Instead of global collapse, another possibility is that as a result of thermal instability, some cold dense clumps will be formed, embedding in the hot phase \citep{y03}. This scenario is perhaps more likely and will be adopted in the present paper. The LHAF solution below and above $\dot{M}_{\rm cr,LHAF}$ are called Type I and II LHAFs, respectively \citep{y01}. We note that the idea of two-phase accretion flow was also explored by other authors \citep{gr88,fr88,k98,w12}, although none of them ever calculated the emitted spectrum.

\subsection{ADAF and Type I LHAF models}
The dynamical equations describing a two-temperature ADAF and Type I LHAF are exactly same, which are summarized below \citep{n97,m97,yqn03},
\begin{eqnarray}
v\frac{{\rm d}v}{{\rm d}R} - \Omega^2 R & = & -\Omega_{\rm K}^2 R
-{1\over \rho} \frac {{\rm d} p}{{\rm d}R}, \label{eq:r}\\
v (\Omega R^2-j)  & = &  - \alpha R {p\over\rho}, \label{eq:ph}\\
q_{\rm adv,i}\equiv\rho v \left(\frac{{\rm d} \varepsilon_{\rm i}}{{\rm d}R}-
{p_{\rm i} \over \rho^2} \frac{{\rm d} \rho}{{\rm d}R} \right)
& = & (1-\delta)q_{\rm vis} - q_{\rm ie}, \label{eq:ei}\\
q_{\rm adv,e}\equiv\rho v \left(\frac{{\rm
d} \varepsilon_{\rm e}}{{\rm d}R}- {p_{\rm e} \over \rho^2}
\frac{{\rm d} \rho}{{\rm d}R}\right) & = & \delta q_{\rm vis} + q_{\rm
ie}-q_{\rm rad}, \label{eq:ee}
\end{eqnarray}
where $j$ is the eigenvalue; $\alpha$ is the viscous parameter; $\varepsilon$ presents the specific internal energy; $q_{\rm vis} (\equiv -\alpha p R {{\rm d}\Omega\over{\rm d} R})$ is the total viscous heating rate. $q_{\rm adv}$, $q_{\rm ie}$ and $q_{\rm rad}$ are the energy advection rate, the energy transfer rate by Coulomb collision between ions and electrons, and the radiative cooling rate, respectively. The radiation includes synchrotron, bremsstrahlung, and inverse Compton \citep{ny95,m97}. Subscripts ``i'' and ``e'' denote quantities of ions and electrons, respectively. All other non-specified quantities are of their usual meanings.

The mass continuity equation is,
\begin{equation}
\mdot (R)=\mdot_0 \left({R\over R_{\rm out}}\right)^s,\label{eq:mdotr}
\end{equation}
where $\mdot(R) \equiv -4\upi RH\rho v$ ($H=c_s/\Omega_{\rm K}$ is the scale height at $R$) is the inflow accretion rate. \citet{ywb12} found that the profile of $\dot{M}(R)$ has two parts. When $R\la 10R_{\rm s}$, $s\approx 0$; but outside $10 R_{\rm s}$, $s\approx 0.5$. In the present paper, we set $s=0.4$ throughout the flow for simplicity.

In order to calculate the synchrotron radiation, the magnetic field strength remains to be specified. This is done through a parameter $\beta \equiv p/p_{\rm mag}$, where magnetic pressure $p_{\rm mag}=B^2/8\pi$ and $p$ is the total (gas+magnetic) pressure. Our calculation of synchrotron and bremsstrahlung emission follows \citet{ny95}. For the Compton scattering part, we only adopt local Compton scattering processes (see Sec.\ \ref{s:dis} for discussions on global Compton scattering).

Once the dynamical structure (density, temperature, etc) of hot accretion flow is determined, we can calculate the spectrum \citep{yqn03}. The bolometric luminosity can then be obtained by integrating over frequency. When outflow is included, the definition of radiative efficiency is a bit subtle since $\dot{M}$ now is a function of $R$. Now we define the radiative efficiency as,
\begin{equation}
\epsilon = L/\mdot_{\rm net}c^2,\label{eq:epsilon2}
\end{equation}
where $\mdot_{\rm net}$ is the accretion rate at the Schwartzschild radius $R_{\rm s}\equiv 2 GM_{\rm BH}/c^2$. This choice adopts the lowest accretion rate, thus gives the highest radiative efficiency. If we define the efficiency using the accretion rate at $R_{\rm out}$, obviously it will be,
\begin{equation}
\epsilon^\prime \equiv {L\over\mdot(R_{\rm out}) c^2}= \left(\frac{R_{\rm s}}{R_{\rm out}}\right)^s\epsilon.\label{eq:epsilon3}
\end{equation}

\begin{figure*}
\centerline{\includegraphics[width=15 cm]{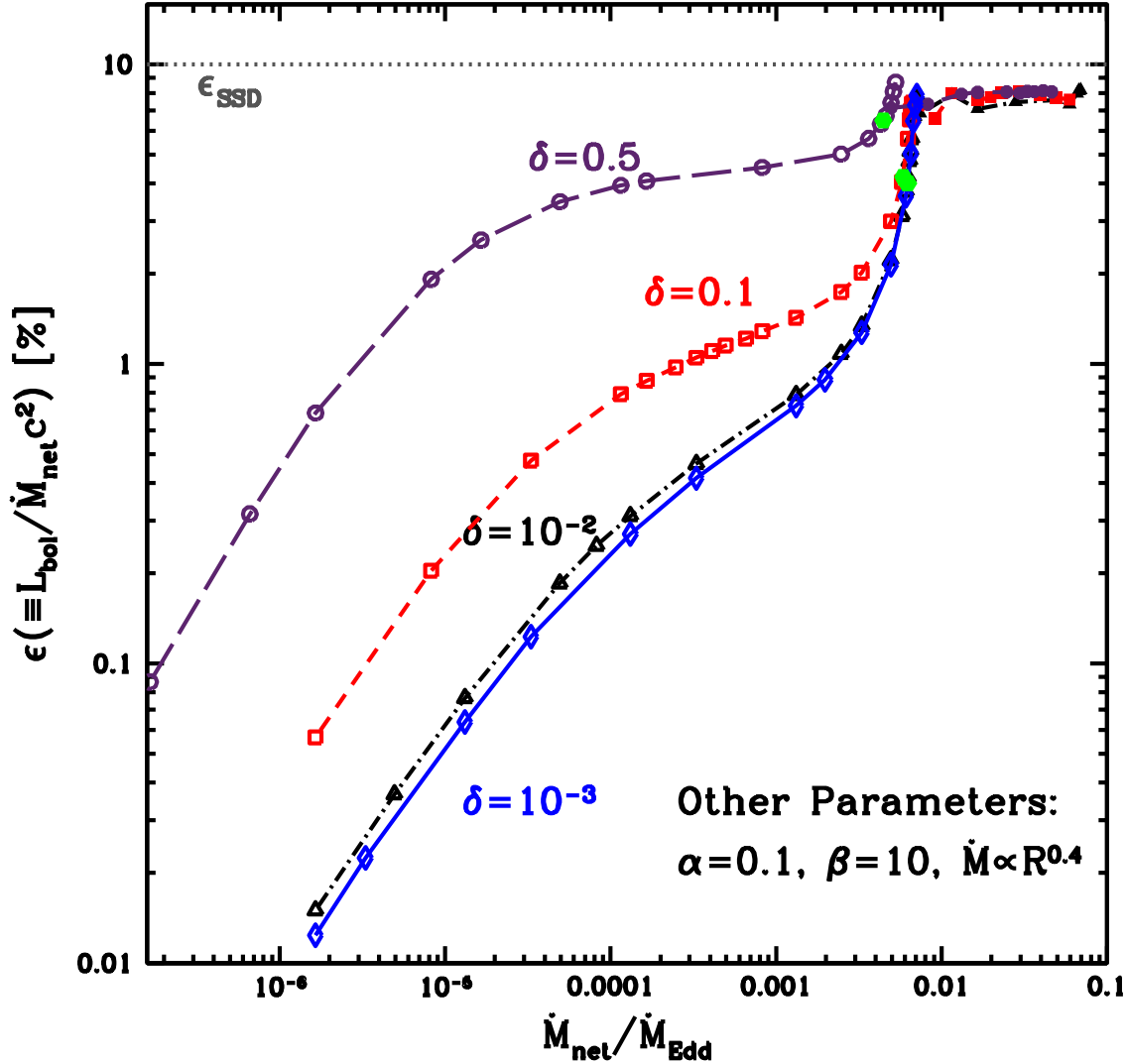}}
\caption{The radiative efficiency of hot accretion flows defined in Eq.\ (\ref{eq:epsilon2}) as a function of the net accretion rate. The open symbols represent results of ADAF or Type I LHAF, while the filled symbols are for two-phase accretion model. The solid (with data shown as diamonds), dot-dashed (with triangles), dashed (with squares), and long-dashed lines (with circles) represent $\delta=10^{-3}$,  $\delta =10^{-2}$, $\delta=0.1$ and $\delta=0.5$, respectively. The two-phase model for $\delta=10^{-3}$ case is similar to $\delta=10^{-2}$ one, thus is not shown. The dotted curve is the radiative efficiency of standard thin disk model ($\epsilon_{\rm SSD} \equiv 0.1$). The filled hexagons mark the value of $\mdot_{\rm cr,ADAF}$ for each choice of $\delta$. Power-law fitting results are presented in Eq.\ (\ref{epsfit}) and Table \ref{tabfit}.}
\label{fig:eff}
\end{figure*}

\subsection{Two-phase accretion model (Type II LHAF)}

As explained above, the accretion flow is likely to have a two-phase structure when $\dot{M}\ga \dot{M}_{\rm cr,LHAF}$ (but lower than another critical accretion rate, $\sim 10\mdot_{\rm cr, LHAF}$, above which hot solutions cease to exist). In this type of accretion flow, cold clumps will emit optical/UV radiation, and these photons will also serve as seed photons for the Comptonization process, which is responsible for X-ray emission. The whole process is obviously very complicated, depending on the detailed dynamics of the two-phases accretion flow such as the filling factor and temperature of clumps and so on. This is beyond the scope of the present paper. Here following \citet{yz04}, we adopt a simplified approach, which is based on the Compton $y$-parameter,
\begin{equation}
y=4\theta_e(1+4\theta_e)(\tau+\tau^2),\label{eq:ypar}
\end{equation}
where $\tau=\sigma_T n_e H$ is the vertical optical depth of the accretion flow. Although $y$-parameter is in general expected to be a function of radius, we now focus on the inner region, $R\la 20 R_{\rm s}$, where most of the radiation comes from. We use $y=constant$ to replace the electron energy equation Eq.\ (\ref{eq:ee}). This implies that we assume the gradient of $y$ does not affect our result significantly in this region. Combined with other dynamical equations, we can calculate the dynamics of the two-phase accretion model for a given $y$-parameter.

The total luminosity is calculated by a different approach. Hold in mind that the radiative cooling rate per unit volume $q_{\rm rad} = F/H$, where $F$ is the radiative flux at the surface of the accretion flow. We determine the bolometric luminosity as,
\begin{equation}
L=2\int 2\pi R FdR=4\pi\int RHq_{\rm rad}dR.
\end{equation}
To determine the quantity $q_{\rm rad}$, we note that at high accretion rate, the advection factor of electrons is nearly zero (refer to the dot-dashed and long-dashed curves in Fig.\ \ref{fig:adv50}). Assuming that this is also true for a two-phase accretion flow, we set $q_{\rm adv,e} = 0$ in Eq.\ (\ref{eq:ee}) and calculate the radiative cooling rate as $q_{\rm rad} = \delta q_{\rm vis}+q_{\rm ie}$.

\section{Numerical Results}
\label{s:result}

\subsection{Radiative efficiency}
We set the black hole mass $\mbh$ to $10 \msun$. For supermassive black holes, we find that the results are  similar. The outer boundary is fixed to be $R_{\rm out}=10^2 R_{\rm s}$. So we have $\dot{M}_{\rm net}=(R_{\rm in}/R_{\rm out})^s\mdot_0=0.16\mdot_0$. Here $\mdot_0\equiv \mdot(R_{\rm out})$. We adopt various $\delta$: $\delta=10^{-3}, 10^{-2}, 0.1$, and $0.5$. Throughout this paper we set $\alpha=0.1$. Numerical simulations show that if the $\alpha$ viscosity is intrinsically the magnetic stress associated with the MHD turbulence driven by magnetorotational instability, as widely accepted, we usually have $\alpha \beta = {\rm constant}$, with the constant being of order unity \citep{bpv08}. We therefore set $\beta=10$.

The results of efficiency are shown in Fig.\ \ref{fig:eff}. For given outflow strength ($s=0.4$), the critical net accretion rates ($\mdot_{\rm cr,ADAF}$, $\mdot_{\rm cr, LHAF}$) are similar for various $\delta$: $(6.3,7.1)\times10^{-3}\medd ~~(\delta = 10^{-3})$, $(6.2,7.1)\times10^{-3}\medd ~~(\delta = 10^{-2})$, $(5.9,6.6)\times10^{-3}\medd ~~(\delta = 0.1)$ and $(4.4,5.3)\times10^{-3}\medd ~~(\delta = 0.5)$, respectively. Several results can be seen from Fig.\ \ref{fig:eff}. \begin{itemize}
\item The radiative efficiency for ADAF is positively correlated with the mass accretion rate, as expected. When $\dot{M}_{\rm net}\la 2\times 10^{-5}\medd$, the slopes for various $\delta$ are similar, the efficiency can be described by  $\epsilon \propto \mdot^{0.7}$. This is flatter than previous estimations of $\epsilon \propto \mdot$ in \citet{nmq98} where $\delta=10^{-3}$. This discrepancy is not because of $\delta$ or outflow effect, but seems to be simply because that the  estimation in previous work is ``rough''. From Figure 7 in \citet{nmq98}, as the accretion rate changes by $3$ orders of magnitude, i.e., from $10^{-4}\medd$ to $10^{-1}\medd$, the bolometric luminosity varies by $5$ orders of magnitude, i.e., from $10^{-7} \ledd$ to $10^{-2} \ledd$. So we should also have $\epsilon \propto \mdot^{0.6-0.7}$, fully consistent with our result.

\item In the ADAF regime, the radiative efficiency strongly depends on the value of $\delta$. This is because a larger $\delta$ implies that more energy will be received by the electrons, subsequently higher radiative efficiency. But note that when $\dot{M}$ is small, the efficiency is still very low. In the case of the accretion flow in Sgr A*, if we adopt the definition of Eq.\ (\ref{eq:epsilon3}), the radiative efficiency will be $\sim 4\times10^{-5}$, which is lower than that of a standard thin disk by a factor of $4\times10^{-4}$ \citep{yqn03}. Since $R_{\rm out}=R_{\rm Bondi}\approx 10^5 R_{\rm s}$ and $s\approx 0.3$ are adopted in \citet{yqn03}, the mass loss in outflow contributes $(1/10^5)^{0.3}\approx 0.04$, the other factor ($4\times10^{-4}/0.04\approx10^{-2}$) is because of energy advection by both ions and electrons (refer to section \ref{sec:eibalance}).

\item When $\dot{M}_{\rm net}\sim \mdot_{\rm cr,ADAF}$, for different $\delta$ the slopes are all very steep and the values of $\epsilon$ become comparable. This is because in this regime of $\dot{M}$, $q_{\rm vis,e}$ is compensated by $q_{\rm ie} (\sim q_{\rm vis,e})$ in the electron energy equation. Moreover, the main radiative process is the Comptonization of synchrotron photons. This process is very sensitive to the optical depth, or accretion rate, of the accretion flow. This is why we have a steep slope.

\item When $\delta=0.5$, the radiative efficiency can be as high as $3\%$, even when $\mdot_{\rm net}$ is as low as $\sim 2\times10^{-5}\medd$. Of course, if the definition of Eq.\ (\ref{eq:epsilon3}) is adopted, the efficiency will be lower by a factor of $(R_{\rm s}/R_{\rm out})^s$ ($\approx 0.16$ in our case). Still, this implies that the efficiency of ADAFs is not as low as people sometimes imagine. For other values of $\delta$, when $\dot{M}_{\rm net}\ga \dot{M}_{\rm cr,ADAF}$, i.e., in LHAFs, the radiative efficiency is all quite high. One application of this result is that we should not observe any large luminosity change during the state transitions from hard to soft in black hole X-ray binaries. This is well consistent with X-ray observations \citep[e.g.,][]{z04}.

\item The radiative efficiency of the two-phase accretion model is nearly independent of the accretion rate, similar to the standard thin disk model. The radiative efficiency can be as high as $\sim 8\%$ in our definition. This is close to or even slightly larger than that of the standard thin disk (note we consider a Schwarzschild black hole). For two-phase accretion flows, the electron advection is zero. Summing up equations (\ref{eq:ei}) and (\ref{eq:ee}), we have $q_{\rm rad}=q_{\rm vis}-q_{\rm adv,i}$. Since both $q_{\rm vis}$ and $q_{\rm adv,i}$ are proportional to density thus $\dot{M}$, $\epsilon$ will be a constant (also independent of $\delta$). As $q_{\rm adv,i}<0$ for a LHAF while $q_{\rm adv,i}=0$ for a SSD, the efficiency of LHAF can be slightly higher than that of a standard thin disk, if the outflow effect is not considered.

\item  The range of accretion rate within which two-phase accretion model exists spans a factor of $8-10$ for our chosen parameters. This range is roughly independent of $\delta$.

\end{itemize}

For the convenience of use, in the following we provide a piecewise power-law fitting to efficiency for ADAF and Type I LHAF. We assume,
\begin{equation}
\epsilon (\mdot_{\rm net}) = \epsilon_0 \left({\mdot_{\rm net}\over\mdot_c}\right)^a,\label{epsfit}
\end{equation}
where the normalization $\mdot_c=10^{-2}\medd=10^{-1}\ledd/c^2$. The fitting results can be found in Table\ \ref{tabfit}. The boundary accretion rates are adjusted after the fitting so that a continuous fitting function is achieved.

So far our numerical calculations and fittings are only for $\alpha=0.1$. In the literature, $\alpha=0.3$ is also widely adopted. We therefore have also calculated the $\alpha=0.3$ and $\delta=0.5,0.1$ and $10^{-3}$ cases. We find that the following formula, with coefficients taken from corresponding $\alpha=0.1$ cases, presents good fit to the cases of $\delta=0.5$ and $0.1$,
\begin{equation}
\epsilon (\mdot_{\rm net}) = \epsilon_0 \left({\alpha\over 0.1}\right)^{0.5}\left({\mdot_{\rm net}\over\mdot_c}\right)^a,
\label{epsfitnew}\end{equation}
here $\mdot_c =0.1 \alpha\medd=\alpha \ledd/c^2$. Note that for $\alpha=0.1$, it recovers the previous definition. For $\delta=10^{-3}$ case, if $\mdot_{\rm net} \la 7\times10^{-2} \alpha^2\medd$ (the typical ADAF regime), the efficiency can be nicely fitted by the following formula,
\begin{equation}
\epsilon (\mdot_{\rm net}) = \epsilon_0 \left({\mdot_{\rm net}\over\alpha^2 \mdot_{\rm Edd}}\right)^a.
\label{epsfitnew2}\end{equation}
When the accretion rate is higher, especially when $\mdot_{\rm net}\ga 5\times10^{-2}\alpha\medd$ (transition from ADAF to Type I LHAF), Eq.\ (\ref{epsfitnew}) presents good fit to the data.

\begin{table} \caption{\large Piecewise power-law fit formulae of radiative efficiencies for ADAF and Type I LHAF.}\label{tabfit}
\begin{tabular}{lcccc}
\hline\hline Cases & $\mdot_{\rm net}/\medd$ Range & $\epsilon_0$  & index $a$\\
\hline
$\delta=0.5$ & $\la 2.9\times10^{-5}$ & $1.58$ & $0.65$\\
             & $2.9\times10^{-5} - 3.3\times10^{-3}$ & 0.055 & 0.076 \\
             & $3.3\times10^{-3} - 5.3\times10^{-3}$ & $0.17$ & $1.12$\\
\hline
$\delta=0.1$ & $\la 9.4\times10^{-5}$ & $0.12$ & $0.59$\\
             & $9.4\times10^{-5} - 5.0\times10^{-3}$ & $0.026$ & $0.27$\\
             & $5.0\times10^{-3} - 6.6\times10^{-3}$ & $0.50$ & $4.53$\\
\hline
$\delta=10^{-2}$ & $\la 1.6\times10^{-5}$ & $0.069$ & $0.69$\\
              & $1.6\times10^{-5} - 5.3\times10^{-3}$ & $0.027$ & $0.54$\\
              & $5.3\times10^{-3} - 7.1\times10^{-3}$ & $0.42$ & $4.85$\\
\hline
$\delta=10^{-3}$ & $\la 7.6\times10^{-5}$ & $0.065$ & $0.71$\\
               & $7.6\times10^{-5} - 4.5\times10^{-3}$ & $0.020$ & $0.47$\\
               & $4.5\times10^{-3} - 7.1\times10^{-3}$ & $0.26$ & $3.67$\\
\hline
\end{tabular}
\begin{quote}
\ Note: See context for the definition of radiative efficiency $\epsilon$. The fitting takes the form $\epsilon (\mdot_{\rm net}) = \epsilon_0 \left(\mdot_{\rm net}/\mdot_c\right)^a$, where the normalization accretion rate is fixed at $\mdot_c=10^{-2}\medd=10^{-1}\ledd/c^2$. For other values of $\alpha$, the fitting formulae can be found in Eq.\ (\ref{epsfitnew}) or Eq.\ (\ref{epsfitnew2}), depending on $\delta$ and accretion rate.
\end{quote}
\medskip
\end{table}

\begin{figure}
\centerline{\includegraphics[width=8.cm]{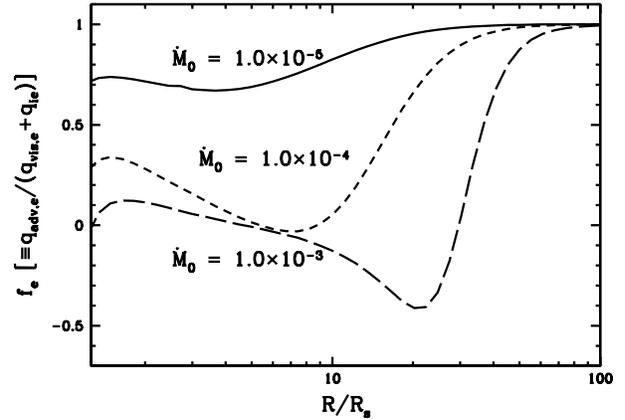}}
\caption{The advection factor of electrons ($f_{\rm e}$) when $\dot{M}_0\la 10^{-3}\medd$ for $\delta = 0.5$. The accretion rate at $R_{\rm out}=10^2 R_{\rm s}$ for each curve is labeled in the plot (in unit of $\medd$).}
\label{fig:adv50_e_low}
\end{figure}

\subsection{Energy balance for ions and electrons}
\label{sec:eibalance}
We now investigate the energy balance relationship among the terms in the energy equations of ions and electrons. We define the advection factors for ions and electrons as follows,
\begin{eqnarray}
f_{\rm i} &=& {q_{\rm adv,i}\over q_{\rm vis,i}}=1-{q_{\rm ie}\over (1-\delta)q_{\rm vis}}, \label{eq:f_i}\\
f_{\rm e} &=& {q_{\rm adv,e}\over q_{\rm vis,e}+q_{\rm ie}}=1-{q_{\rm rad}\over \delta q_{\rm vis}+q_{\rm ie}}\label{eq:f_e}.
\end{eqnarray}
Note that from Eq.\ (\ref{eq:f_i}), one can get,
\begin{equation}
{q_{\rm vis,e}\over q_{\rm ie}} = {\delta\over (1-\delta)(1-f_{\rm i})},
\end{equation}
which characterizes the relative importance of $q_{\rm vis,e}$ and $q_{\rm ie}$ in the energy equation of electrons. Evidently from this equation, we find that for large
value of $\delta$ (e.g., $\delta\ga0.1$), viscous heating to electrons will be the main heating term to the electrons, provided that the ions are advection-dominated (i.e., $f_{\rm i}\ga 0.9$). Even for lower $\delta$, $q_{\rm vis,e}$ will still be the main heating term at low accretion rate, where $f_{\rm i}=1$.

\begin{figure}
\centerline{\includegraphics[width=8.cm]{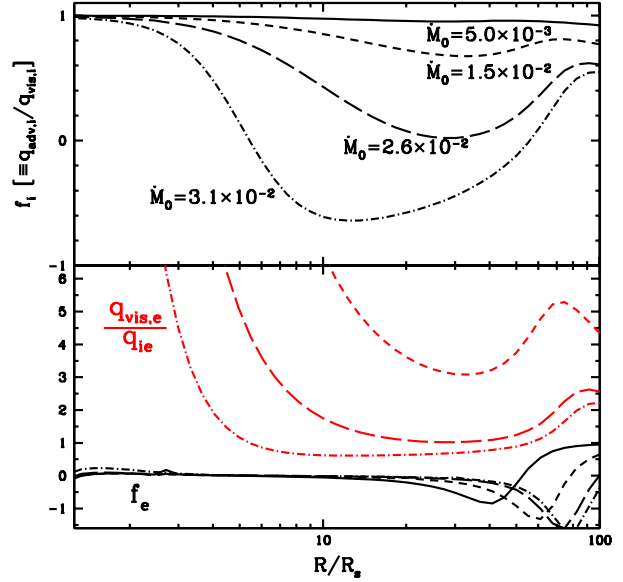}}
\caption{The energy balance relationship for ions and electrons when $\dot{M}_0\ga 5\times 10^{-3}\medd$ for $\delta = 0.5$. {\it Upper panel:} advection factor of ions $f_i$. {\it Lower panel:} advection factor of electrons $f_{\rm e}$ (black) and $q_{\rm vis,e}/q_{\rm ie}$ (red). The  lines with the same type have the same accretion rate with the upper panel. The curve of $q_{\rm vis,e}/q_{\rm ie}$ for $\mdot_0=5.0\times10^{-3}\medd$ is not shown here since its value is too large. }
\label{fig:adv50}
\end{figure}

Below we focus only on the case of $\delta=0.5$ (the long-dashed curve in Fig.\ \ref{fig:eff}), since this value of $\delta$ is most favored theoretically from the detailed study of Sgr A* \citep{yqn03}. The results are shown in Figs.\ \ref{fig:adv50_e_low}\&\ref{fig:adv50}. Note that $q_{\rm vis,i}=q_{\rm vis,e}$ for this choice of $\delta$. We describe the results at the following three regimes of accretion rate,\\

\begin{itemize}
\item $\mdot_0\la 3.0\times10^{-5}\medd$. \\
   In this regime, both the ions and electrons are strongly advection dominated, i.e., $f_{\rm i}=1, f_{\rm e}=1$. Both the Coulomb coupling and the radiative cooling rate are negligible compared to the viscous heating rate. Specifically, we have,
   \begin{eqnarray}
   {\rm ~~~~~~~~~ions:}& &q_{\rm adv,i} \approx q_{\rm vis,i}\gg q_{\rm ie},\nonumber\\
   {\rm ~~~~electrons:}& &q_{\rm adv,e} \approx q_{\rm vis,e}\gg q_{\rm rad} \& q_{\rm ie}.
   \end{eqnarray}

\item $3.0\times10^{-5}\medd\la\mdot_0\la1.0\times10^{-2}\medd$. \\
    In this regime, the accretion rate is still low enough, the ions remains advection-dominated ($f_{\rm i}\approx1$). For the electrons, the radiative cooling rate is high because of the high density of the accretion flow. As illustrated in Figs.\ \ref{fig:adv50_e_low}\&\ref{fig:adv50}, the electrons are radiation-dominated, i.e. the advection factor $f_{\rm e}\approx 0$ in the inner $R<20 R_{\rm s}$ regions.
   \begin{eqnarray}
   {\rm ~~~~~~~~~ions:}& &q_{\rm adv,i} \approx q_{\rm vis,i}\gg q_{\rm ie},\nonumber\\
   {\rm ~~~~electrons:}& &q_{\rm rad} \approx q_{\rm vis,e}\gg q_{\rm adv,e}\&q_{\rm ie}.
   \end{eqnarray}

\item $1.0\times10^{-2}\medd\la\mdot_0\la3.2\times10^{-2}\medd$. \\
   This is the transition regime from an ADAF to a Type I LHAF. As the accretion rate (or the density) increases, the Coulomb coupling between ions and electrons becomes so strong that the ions are no longer advection-dominated, i.e. $q_{\rm ie}\sim q_{\rm vis,i}$. Above $\mdot_{\rm cr, ADAF}$, we have $q_{\rm ie}\ge q_{\rm vis,i}$ in some regions of the hot flow, which means that the flow enters into the Type I LHAF regime. All the energy terms in the ion energy equation are roughly comparable to each other. The electrons radiate away nearly all the energy they receive via viscous heating and Coulomb collision, i.e. $f_{\rm e}=0$. The relative importance of $q_{\rm vis,e}$ and $q_{\rm ie}$ depends on various parameters, i.e. $\mdot_0$ and $\delta$, and the radius $R$.
   \begin{eqnarray}
  {\rm ~~~~~~~~~ions:}& &q_{\rm adv,i} \sim q_{\rm vis,i}\sim q_{\rm ie},\nonumber\\
   {\rm ~~~~electrons:}& &q_{\rm rad} \approx (q_{\rm vis,e}+q_{\rm ie})\gg q_{\rm adv,e}.
   \end{eqnarray}
\end{itemize}

For other choices of $\alpha$, the above arguments still hold, except that the accretion rate regime is replaced by $(\alpha/0.1)$ times the value listed above.

Above $\mdot_{\rm cr,LHAF}$, purely hot solutions do not exist; and the accretion flow enters the two-phase regime. In this regime, we expect that $f_{\rm e}=0$, i.e. $q_{\rm rad}=q_{\rm vis,e}+q_{\rm ie}$.

\section{discussions: caveats and radio-X-ray correlation}
\label{s:dis}

In our calculations, we only consider the local Compton scattering, namely the scattering between photons and electrons occurred at the same region where the photons are produced. However, since a hot accretion flow is usually optically thin in the radial direction, the photons produced at one certain radius can in principle travel for a long distance and collide with electrons at another radius. Such a ``global'' Compton scattering effect has been systematically investigated in previous works \citep{po01,po07,yxo09,x10,nxz12}. It was found that it plays a significant cooling and heating roles in the region of $50 R_{\rm s}\la R \la 100 R_{\rm s}$ and $R\ga 5\times 10^3 R_{\rm s}$, respectively, when the accretion rate is high enough so that the total luminosity emitted from the accretion flow $L_{\rm bol}\ga 2\%\ledd$ \citep{yxo09,x10}. One consequence is that the radiative efficiency will be lower by a factor of 2 compared to the case that this effect is not taken into account \citep{x10}. In addition, the highest luminosity a hot accretion flow can emit will be constrained to be $L_{\rm bol}\la 1\%\ledd$. Above this limit, the global Compton cooling and heating will be so strong that no steady hot solution can be found and the system will ``oscillate'' \citep{yxo09}. Note that if $L_{\rm bol} \la 2\%\ledd$, or outer boundary radius of hot flow is small, i.e. $R_{\rm out}\la 50-100R_{\rm s}$, the global Compton scattering effects will be unimportant. The latter is the case of luminous hard state of black hole X-ray binaries.

In our model, we assume that the magnetic filed is tangled and weak, thus it does not play any dynamical role. Numerical simulations have shown that a large-scale toroidal magnetic field is likely to exist in the inner region of the accretion flow, imposed on the stochastic component \citep[e.g.][]{hkdh04}. The effect of such a field has been studied by self-similar approaches \citep{af06,agn08,b09} or global calculations \citep{o07,o12}. Especially, the global solution with strong large-scale magnetic fields indicates an increase in the highest luminosity a hot accretion flow can achieve \citep{o12}.

Throughout this paper, we fix the outer boundary condition ($T_{\rm i}, T_{\rm e}, v$) in our calculations. The effect of outer boundary condition on the dynamics of accretion flow has been studied in \citet{y00}. Obviously, it will also influence the radiative efficiency.


A correlation between the radio and the 2-10 keV luminosity ($L_{\rm X}$) has been found among the hard state of black hole X-ray binaries and low-luminosity active galactic nuclei \citep{c00,c03,g03,mhd03,gmf12,c12}, which is well described by a power-law, $L_{\rm Radio} \propto L_{\rm X}^p$ with index $p \sim 0.6$. Recently \citet{z11} found that this correlation extends to intermediate and soft states for Cyg X-1, if only the luminosity from hot disc is used. This correlation has been quantitatively explained by the coupled jet-ADAF model in \citet{yc05}, in which the radio and X-ray emissions are dominated by the radiation from the jet and ADAF, respectively.
It is interesting to note that there are now growing number of sources which show that when $L_{\rm X}\ga 4\times 10^{36}\ergs$ the radio/X-ray correlation follows a steeper power-law, with index $p \sim 0.98$ or $1.4$ \citep{c11,gmf12,c12}. Below this critical luminosity, the sources return to the $\sim 0.6$ correlation at $\sim 10^{35}\ergs$. Between $4\times 10^{36}\ergs$ and $10^{35}\ergs$, the radio luminosity remains almost unchanged. \citet{c11} proposed that one way to explain the steep ($p \sim 1.4$) correlation is that the radiative efficiency of the hot accretion flow is independent of the accretion rate, if the ratio of the mass loss rate in the jet and the accretion rate in the accretion flow is a constant. As shown by Fig. 1, this is the case for our two-phase accretion flow (Type II LHAF). Moreover, as also shown by this figure, the efficiency curve of Type I LHAF is very steep, which means that a small change of accretion rate will result in a large change of $L_{\rm X}$. This feature is obviously attractive to explain the ``flat transition'' between $4\times 10^{36}\ergs$ and $10^{35}\ergs$. The reason why some sources follow a single $\sim 0.6$ correlation while others follow three branches is simply because of different values of $\alpha$ among these sources. If $\alpha$ is large, $\dot{M}_{\rm cr,ADAF}$ will be large, so no transition to LHAFs will occur throughout the evolution of $\dot{M}$ during the outburst. This is why we can only observe one single $\sim 0.6$ correlation. If on the other hand $\alpha$ in a source is small, $\dot{M}_{\rm cr,ADAF}$ will be small thus the sources will enter the two-phase LHAFs regime during the outburst. In this case, three branches of correlation should be expected. In a future work we plan to investigate the correlation in detail.

\section*{Acknowledgments}

We are grateful to St\'ephane Corbel, Chris Done, Andrzej Zdziarski, and the referee for their valuable comments. This work was supported in part by the Natural Science Foundation of China (grants 10833002, 10825314, 11103059, 11121062, 11133005, and 11203057), the National Basic Research Program of China (973 Program 2009CB824800), and the CAS/SAFEA International Partnership Program for Creative Research Teams. FGX was also supported in part by a project No. ZDB201204.

\label{lastpage}

\end{document}